\documentclass{article}
\usepackage{url,overcite}
\usepackage[colorlinks,citecolor=blue,urlcolor=blue]{hyperref}
\usepackage{graphicx}
\usepackage{braket}
\usepackage{amsmath}
\usepackage{subfigure}
\usepackage{geometry}
\begin{document}

\markboth{ Manosh T M, Muhammed Ashefas \& Ramesh Babu Thayyullathil}{Effects of Kerr medium in coupled cavities on quantum state transfer}

\title{Effects of Kerr medium in coupled cavities on quantum state transfer }

\author{Manosh T.M., Muhammed Ashefas C.H. and Ramesh Babu Thayyullathil\\ Department of Physics\\
	Cochin University of Science and Technology, India - 682022\\
	$^{\dagger}$manosh@cusat.ac.in}
\date{ }
\maketitle

\begin{abstract}
We study the effect  of  Kerr type nonlinear medium in quantum state transfer. We have investigated the effect of different coupling schemes and Kerr medium  parameters $p$ and $\omega_{{K}}$. We found that, the Kerr medium introduced in the connection channel can act like a controller for quantum state transfer. The numerical simulations are performed without taking the adiabatic approximation. Rotating  wave approximation is used in the atom-cavity interaction  only in the lower coupling regime.
\end{abstract}

\section{Introduction}

Quantum information processing (QIP) need an effective implementation of quantum state transferring schemes. The implementation of such systems with optical cavity has been a state of art. The simplest quantum description of cavity system are well described in the literature\cite{1443594,2012arXiv1203.2410H,PhysRevA.50.1785}. Modification on the Jaynes Cummings model (JCM) has been an active field of research ever since. The two level system (TLS) was then extended to multilevel, multi-cavity, multi atom TLS, \cite{0954-8998-3-2-005,Seke:85,Seke:92}etc.

Optical cavities are very good  candidate  for quantum information processing (QIP)\cite{2004quant.ph..5030C}.  The   nonlinear optical behavior due to  $\chi^{(2)}$, Kerr nonlinearity ($\chi^{(3)}$) etc. has also been used to modify these optical cavities  and are extensively studied  both theoretically  as well as experimentally \cite{scully_zubairy_1997,10.1007/978-3-540-68484-8}. Implementation of such optical cavities and its use  in QIP has been an area of active research since late 1960s \cite{PhysRev.155.980,Puri2017}. State transfer in coupled cavities appears to be a reliable platform for data transfer  in QIP \cite{2004quant.ph..5030C,Man2015,PhysRevA.75.022310,PhysRevA.75.032331,Nohama2007,PhysRevA.93.062339}. Here we discuss  coupled cavity systems with and without Kerr type nonlinearity.

In this paper we discuss the quantum state transfer (QST) in a linearly coupled cavity array with and without Kerr medium. We introduce the Kerr medium in the connection channel alone. We found that, the presence of a Kerr medium can be used as a controller in the QST. 
\section{Linearly coupled cavities}

\begin{figure}
	\begin{center}
		\includegraphics[width=8cm,height=3cm]{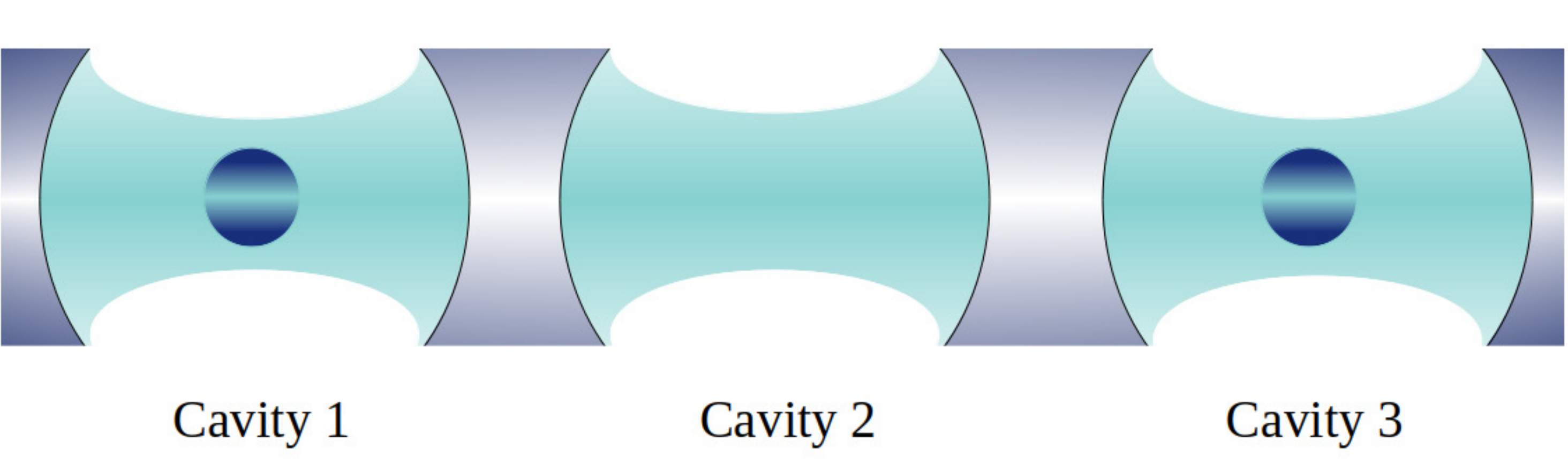}
		\caption{Three linearly coupled cavity with one qubit at either ends.}
		\label{figure1}
	\end{center}
\end{figure}

A quantum  state carries an information, which has to be transfered from one place to another without any loss or modification. As quantum mechanics follows \textit{no cloning} theorem\cite{Wootters1982}, it is not possible to send an exact copy of the information\cite{Park1970}. Here we consider a quantum state in the cavity 1 which has to be transfered to a cavity 3, through an  intermediate cavity 2. The system is illustrated in the figure (\ref{figure1}).

Here in all the 3 cavities we have single mode photon field with a two level atom  (or qubit) in  1st and 3rd  cavities. The system can be described  by the Hamiltonian,	 
\begin{equation}
\hat{H}=\hat{H}_0+\hat{H}_I, 
\label{equation1}
\end{equation}
\noindent where, $\hat{H}_0$ is the free Hamiltonian and $\hat{H}_I$  is the interaction Hamiltonian and we have 
\begin{equation}
\hat{H}_0=\hbar\omega_ca^{\dagger}_1a_1+\hbar\omega_ca^{\dagger}_2a_2+\hbar\omega_ca^{\dagger}_3a_3+\frac{\hbar}{2}\omega_a\sigma_z^{(1)}+\frac{\hbar}{2}\omega_a\sigma_z^{(3)},
\end{equation}
\begin{equation}
\begin{split} 
\hat{H}_I=&\lambda_1\left(a^{\dagger}_1\sigma_-^{(1)}+a_1\sigma_+^{(1)}\right)+\lambda_3\left(a^{\dagger}_3\sigma_-^{(3)}+a_3\sigma_+^{(3)}\right)\\ 
&+J_{12}\left(a_1^{\dagger}a_2+a_1a_2^{\dagger}\right)+J_{23}\left(a_2^{\dagger}a_3+a_2a_3^{\dagger}\right),
\end{split}
\end{equation}
\noindent Here $\lambda_i\ (i=1,3)$ are the atom field coupling constant,  $J_{lm}$ are the coupling strength between the cavities $l$ and $m$, $a_i\ (i=1,2,3)$ denotes the field annihilation operator and $\sigma_z^{i},\sigma_+^{i}$ and $\sigma_-^{i}\ (i=1,3)$ are the atomic  operators for the $i$th cavity. A tensor product state of the system can be written as,
\begin{equation}
\ket{\psi}=\ket{k_1n_1n_2k_3n_3},
\label{equation4}
\end{equation}
where $k_i=0$ and $k_i=1$ correspond to ground and excited state respectively of the atom (qubit) in the $i$th cavity and $n_i$ represents the number of photons in the $i$th cavity. Thus, if we 
consider a state   with maximum of one excitation at a time, the corresponding general state may be written as
\begin{equation}
\begin{split}
\ket{\psi(t)}=& q_1(t)\ket{10000}+f_1(t)\ket{01000}+f_2(t)\ket{00100}\\&+q_3(t)\ket{00010}+f_3(t)\ket{00001},
\end{split}
\end{equation}
where $q_i(t)$ and $f_i(t)$ respectively are the atomic and field excitation coefficients in the $i$th cavity.
The dynamics of the system can now be studied by solving the corresponding  Schr\"{o}dinger equation. For convenience,  we can take the atomic transition frequency, $\omega_a$ and the field frequency,  $\omega_c$ as the same. Thus the detuning, $\Delta=\omega_a-\omega_c=0$ and we denote,  $\omega_a=\omega_c=\omega$. 
The state vector in the interaction picture,
\begin{equation}
\ket{\psi(t)}_I=e^{\frac{i}{\hbar}\hat{H}_0t}\ket{\psi(t)}_S,
\end{equation}
satisfies the evolution equation,
\begin{equation}
\hat{\mathcal{H}}_I\ket{\psi(t)}_I=i\hbar\frac{\partial}{\partial t}\ket{\psi(t)}_I,
\end{equation}
where we have 
\begin{equation}
\hat{\mathcal{H}}_I=e^{\frac{i}{\hbar}\hat{H}_0t}\hat{H}_Ie^{-\frac{i}{\hbar}\hat{H}_0t}=\hat{H}_I,
\end{equation}
since  $\left[\hat{H}_0,\hat{H}_I\right]=0$ because the  detuning is set as zero.
The differential equations for $q_i(t)$ and $f_i(t)$ can be obtained  as, 

\begin{align}
i\hbar\frac{\partial}{\partial t}q_1(t) &= \lambda_1f_1(t),
\label{equation9a}\\
i\hbar\frac{\partial}{\partial t}f_1(t) &= \lambda_1q_1(t)+J_{12}f_2(t),\label{equation9b}\\
i\hbar\frac{\partial}{\partial t}f_2(t) &= J_{12}f_1(t)+J_{23}f_3(t),\label{equation9c}\\
i\hbar\frac{\partial}{\partial t}q_3(t) &= \lambda_3f_3(t),\label{equation9d}\\
i\hbar\frac{\partial}{\partial t}f_3(t) &= \lambda_3q_3(t)+J_{23}f_2(t).
\label{equation9e}
\end{align}

\noindent These equations can be  solved numerically  and we can investigate  how the coupling parameters affects the quantum state transfer. 	

\subsection{Analytical approach}
The Laplace transform of the equations (\ref{equation9a}) to (\ref{equation9e}) can be written as, ($\hbar=1$)

\begin{align}
i\left[q_1(0)+sQ_1(s)\right] &= \lambda_1F_1(s)\label{equation10a},\\
i\left[f_1(0)+sF_1(s)\right] &= \lambda_1Q_1(s)+J_{12}F_2(s)\label{equation10b},\\
i\left[f_2(0)+sF_2(s)\right] &= J_{12}F_1(s)+J_{23}F_3(s)\label{equation10c},\\
i\left[q_3(0)+sQ_3(s)\right] &= \lambda_3F_3(s)\label{equation10d},\\
i\left[f_3(0)+sF_3(s)\right] &= \lambda_3Q_3(s)+J_{23}F_2(s).
\label{equation10e}
\end{align}

\noindent For further simplicity we can take the value of $\lambda_1=\lambda_3=\lambda$ and $J_{12}=J_{23}=J$. Now solving the equations (\ref{equation10a}) to (\ref{equation10e})   for an initial condition, $q_1(0)=1$ and $q_2(0)=f_1(0)=f_2(0)=f_3(0)=0$, results in,

\begin{align}
Q_1(s) &=- \frac{J^{2} \lambda^{2} + 2.0 J^{2} s^{2} + \lambda^{2} s^{2} + s^{4}}{s \left(2.0 J^{2} \lambda^{2} + 2.0 J^{2} s^{2} + \lambda^{4} + 2.0 \lambda^{2} s^{2} + s^{4}\right)}\label{equation11a},\\
F_1(s) &= \frac{i \lambda \left(J^{2} + \lambda^{2} + s^{2}\right)}{2.0 J^{2} \lambda^{2} + 2.0 J^{2} s^{2} + \lambda^{4} + 2.0 \lambda^{2} s^{2} + s^{4}}\label{equation11b},\\
F_2(s) &= \frac{J \lambda}{s \left(2.0 J^{2} + \lambda^{2} + s^{2}\right)}\label{equation11c},\\
Q_3(s) &= - \frac{J^{2} \lambda^{2}}{s \left(2.0 J^{2} \lambda^{2} + 2.0 J^{2} s^{2} + \lambda^{4} + 2.0 \lambda^{2} s^{2} + s^{4}\right)}\label{equation11d},\\
F_3(s) &= - \frac{i J^{2} \lambda}{2.0 J^{2} \lambda^{2} + 2.0 J^{2} s^{2} + \lambda^{4} + 2.0 \lambda^{2} s^{2} + s^{4}}.
\label{equation11e}
\end{align}

\noindent Now taking the inverse Laplace transform of equations (\ref{equation11a}) to (\ref{equation11e}), we obtain,

\begin{align}
q_1(t)&=-\frac{1}{2}\left( \cos (\lambda  t)+\frac{\lambda ^2 \cosh \left(t \sqrt{-2 J^2-\lambda ^2}\right)+2 J^2}{4 J^2+2\lambda ^2}\right)\label{equation12a},&\\
q_3(t)&=\frac{1}{2} \left(\cos (\lambda  t)-\frac{\lambda ^2 \cosh \left(t \sqrt{-2
		J^2-\lambda ^2}\right)+2 J^2}{2 J^2+\lambda ^2}\right)\label{equation12b},&\\
f_1(t)&=\frac{1}{2} i \left(\sin (\lambda  t)+\frac{\lambda  \sinh \left(t \sqrt{-2 J^2-\lambda
		^2}\right)}{\sqrt{-2 J^2-\lambda ^2}}\right)\label{equation12c},&\\
f_3(t)&=-\frac{1}{2} i \left(\sin (\lambda  t)-\frac{\lambda  \sinh \left(t \sqrt{-2
		J^2-\lambda ^2}\right)}{\sqrt{-2 J^2-\lambda ^2}}\right)\label{equation12d},&\\
f_2(t)&=-\frac{J \lambda  \left(\cosh \left(t \sqrt{-2 J^2-\lambda ^2}\right)-1\right)}{2
	J^2+\lambda ^2}\label{equation12e}.&	
\end{align}

\noindent The probability can now be calculated as

\begin{equation}
\begin{split}
\left|q_1(t)\right|^2=&\frac{1}{4}\left| \frac{\left(2 J^2+\left(2 J^2+\lambda ^2\right) \cos (t
	\lambda )+\lambda ^2 \cosh \left(t \sqrt{-2 J^2-\lambda ^2}\right)\right) }{\left(2 J^2+\lambda ^2\right) \left(2
	\left(J^*\right)^2+\left(\lambda ^*\right)^2\right)}\right|\times \\
&\left| \frac{\left(2
	\left(\cos (t \lambda )^*+1\right) \left(J^*\right)^2+\left(\lambda ^*\right)^2
	\left(\cos (t \lambda )^*+\cosh \left(\left(t \sqrt{-2 J^2-\lambda
		^2}\right)^*\right)\right)\right)}{\left(2 J^2+\lambda ^2\right) \left(2
	\left(J^*\right)^2+\left(\lambda ^*\right)^2\right)}\right| 
\label{equation13}.
\end{split}
\end{equation}

\begin{align}
\left|q_3(t)\right|^2&=\frac{1}{4}\left| \frac{\left(2 J^2-\left(2 J^2+\lambda ^2\right) \cos (t
	\lambda )+\lambda ^2 \cosh \left(t \sqrt{-2 J^2-\lambda
		^2}\right)\right)^2}{\left(2 J^2+\lambda ^2\right)^2}\right|
\label{equation14}.\\
\left|f_1(t)\right|^2&=
\frac{1}{4}\left| \frac{\left(\sqrt{-2 J^2-\lambda ^2} \sin (t \lambda
	)+\lambda  \sinh \left(t \sqrt{-2 J^2-\lambda ^2}\right)\right)^2}{2 J^2+\lambda
	^2}\right|
\label{equation15}.\\
\left|f_2(t)\right|^2&=\frac{1}{4} e^{-2 \Re\left(t \sqrt{-2 J^2-\lambda ^2}\right)}\left|
\frac{\left(-1+e^{t \sqrt{-2 J^2-\lambda ^2}}\right)^2 \left(-1+e^{\left(t
		\sqrt{-2 J^2-\lambda ^2}\right)^*}\right)^2 J^2 \lambda ^2}{\left(2 J^2+\lambda
	^2\right) \left(2 \left(J^*\right)^2+\left(\lambda
	^*\right)^2\right)}\right|
\label{equation16}.\\
\left|f_3(t)\right|^2&=\frac{1}{16} \left| 2 \sin (t \lambda )-\frac{2 \lambda  \sinh \left(t \sqrt{-2
		J^2-\lambda ^2}\right)}{\sqrt{-2 J^2-\lambda ^2}}\right| ^2
\label{equation17}.
\end{align}

\noindent using equation (\ref{equation13}) to (\ref{equation17}), we can define the population inversion of each qubit as, 

\begin{equation}
\braket{\sigma_z^{(1)}}=\left|q_1(t)\right|^2-\left(\left|q_3(t)\right|^2+\left|f_1(t)\right|^2+\left|f_2(t)\right|^2+\left|f_3(t)\right|^2\right)
\label{equation18},
\end{equation}

\begin{equation}
\braket{\sigma_z^{(3)}}=\left|q_3(t)\right|^2-\left(\left|q_1(t)\right|^2+\left|f_1(t)\right|^2+\left|f_2(t)\right|^2+\left|f_3(t)\right|^2\right)
\label{equation19}.
\end{equation}


\section{Linearly coupled cavities with Kerr medium}

\begin{figure}
	\centering
	\includegraphics[width=8cm,height=3cm]{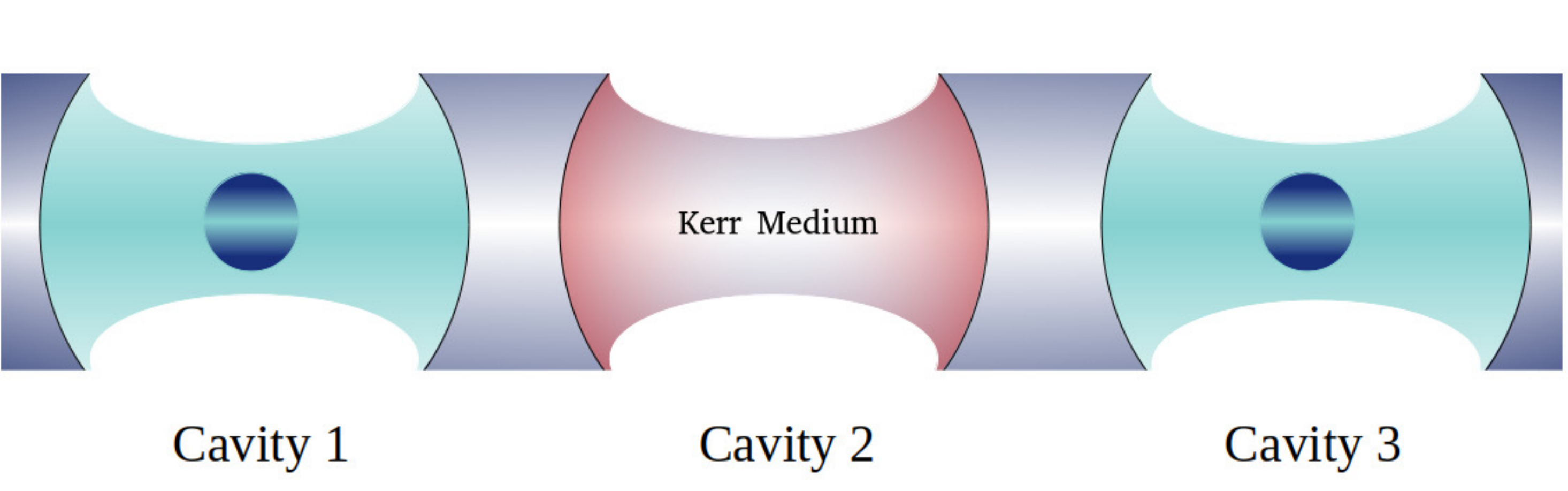}
	\caption{3 linearly coupled cavity with one qubit at either ends and a Kerr medium in the second cavity}
	\label{figure2}
\end{figure}

Nonlinear effects in optical cavities has been studied in the literature \cite{10.1007/978-3-540-68484-8,PhysRev.155.980}. This nonlinear effects can be used to construct effective transfer mechanism in quantum engineering\cite{PhysRevA.85.023826,Puri2017}. Here we consider the 3rd  order nonlinearity, widely known as \textit{Kerr nonlinearity} and we introduce such a nonlinear effect  in the second cavity of the previously described system shown in the figure (\ref{figure1}) and we get the modified system as in figure (\ref{figure2}).
Hamiltonian with a Kerr type nonlinear medium in the second cavity is described by \cite{doi:10.1080/09500349114550961,PhysRevA.39.2969}
\begin{equation}
\hat{H}_{{K}}=\omega_{{K}}b^{\dagger}b+qb^{\dagger2}b^2+p\left(a_2^{\dagger}b+a_2b^{\dagger}\right).
\end{equation}
\noindent where $b$ is  the annihilation operator  of the Kerr medium, $\omega_{{K}}$ denotes  the anharmonic Kerr field frequency, $q$ is the anharmonicity parameter and $p$ represents the field-Kerr medium coupling strength. In the adiabatic limit  the field frequency $\omega$ and medium frequency $\omega_{{K}}$ are very different. 
Now the  Hamiltonian of the new system takes the form, 
\begin{equation}
\hat{\mathcal{H}}=\hat{H}_0+\hat{H}_I+\hat{H}_{{K}}=\hat{\mathcal{H}}_0+\hat{\mathcal{H}}_I.
\label{equation21}
\end{equation}

\noindent where the new free and interaction Hamiltonians are respectively  given as,
\begin{equation}
\hat{\mathcal{H}}_0=\hat{H}_0+\omega_{{K}}b^{\dagger}b+qb^{\dagger 2}b^2,
\end{equation}
\begin{equation}
\hat{\mathcal{H}}_I=\hat{H}_I+p\left(a_2^{\dagger}b+a_2b^{\dagger}\right).
\end{equation}

\noindent With the Kerr medium operator $b$ and $b^{\dagger}$, we need to extend the Hilbert space of states given in equation(\ref{equation4}) and it get modified to to accommodate the new state, which can be defined as,
\begin{equation}
\ket{\psi}=\ket{k_1n_1n_2k_3n_3n_b}.
\label{equation24}
\end{equation}

\noindent where $n_b$ is the bosonic number of the Kerr medium. Here also we can find the Hamiltonian in the interaction picture and we can show that 
\begin{equation}
\left[\hat{\mathcal{H}}_0,\hat{\mathcal{H}}_I\right]=0.
\end{equation}

\noindent The dynamics can be studied by obtaining the differential equations similar to equations (\ref{equation9a}) to (\ref{equation9e}) and solving it. 

\subsection{Analytical approach}

If we take only a maximum of one excitation in the cavity we may write the general state as 

\begin{equation}
\begin{split}
\ket{\Psi(t)}=& q_1(t)\ket{100000}+f_1(t)\ket{010000}+f_2(t)\ket{001000}\\&+q_3(t)\ket{000100}+f_3(t)\ket{000010}+k(t)\ket{000001}.
\end{split} 
\end{equation}

\noindent The differential equations for $q_i(t)$, $f_i(t)$ and $k(t)$ can be obtained  as,

\begin{align}
i\hbar\frac{\partial}{\partial t}q_1(t) &= \lambda_1f_1(t)\label{equation27a},\\
i\hbar\frac{\partial}{\partial t}f_1(t) &= \lambda_1q_1(t)+J_{12}f_2(t)\label{equation27b},\\
i\hbar\frac{\partial}{\partial t}f_2(t) &= J_{12}f_1(t)+J_{23}f_3(t)+pk(t)\label{equation27c},\\
i\hbar\frac{\partial}{\partial t}q_3(t) &= \lambda_3f_3(t)\label{equation27d},\\
i\hbar\frac{\partial}{\partial t}f_3(t) &= \lambda_3q_3(t)+J_{23}f_2(t)\label{equation27e},\\
i\hbar\frac{\partial}{\partial t}k(t) &=\left(\omega_k-\omega-q\right)k(t)+pf_2(t)\label{equation27f}.
\end{align}

\noindent The Laplace transform of equations (\ref{equation27a}) to (\ref{equation27f}) are, ($\hbar=1$)

\begin{align}
i\left[q_1(0)+sQ_1(s)\right] &= \lambda_1F_1(s)\label{equation28a},\\
i\left[f_1(0)+sF_1(s)\right] &= \lambda_1Q_1(s)+J_{12}F_2(s)\label{equation28b},\\
i\left[f_2(0)+sF_2(s)\right] &= J_{12}F_1(s)+J_{23}F_3(s)+pK(s)\label{equation28c},\\
i\left[q_3(0)+sQ_3(s)\right] &= \lambda_3F_3(s)\label{equation28d},\\
i\left[f_3(0)+sF_3(s)\right] &= \lambda_3Q_3(s)+J_{23}F_2(s)\label{equation28e},\\
i\left[k(0)+sK(s)\right] &= \left(\omega_k-\omega-q\right)K(s)+pF_2(s)\label{equation28f}.
\end{align}

\noindent For Kerr medium analytical solutions of can be more rigorous to handle. 
\noindent We do not take the adiabatic approximation even when the $\omega_{{K}}$ is far away from $\omega$ and we investigated the evolution of the system numerically. \cite{JOHANSSON20131234}


\section{Results and discussion}
\subsection{Quantum state transfer without Kerr medium}
\begin{figure}
	\centering     
	\subfigure[]{\label{figure3_3a}\includegraphics[width=0.45\textwidth]{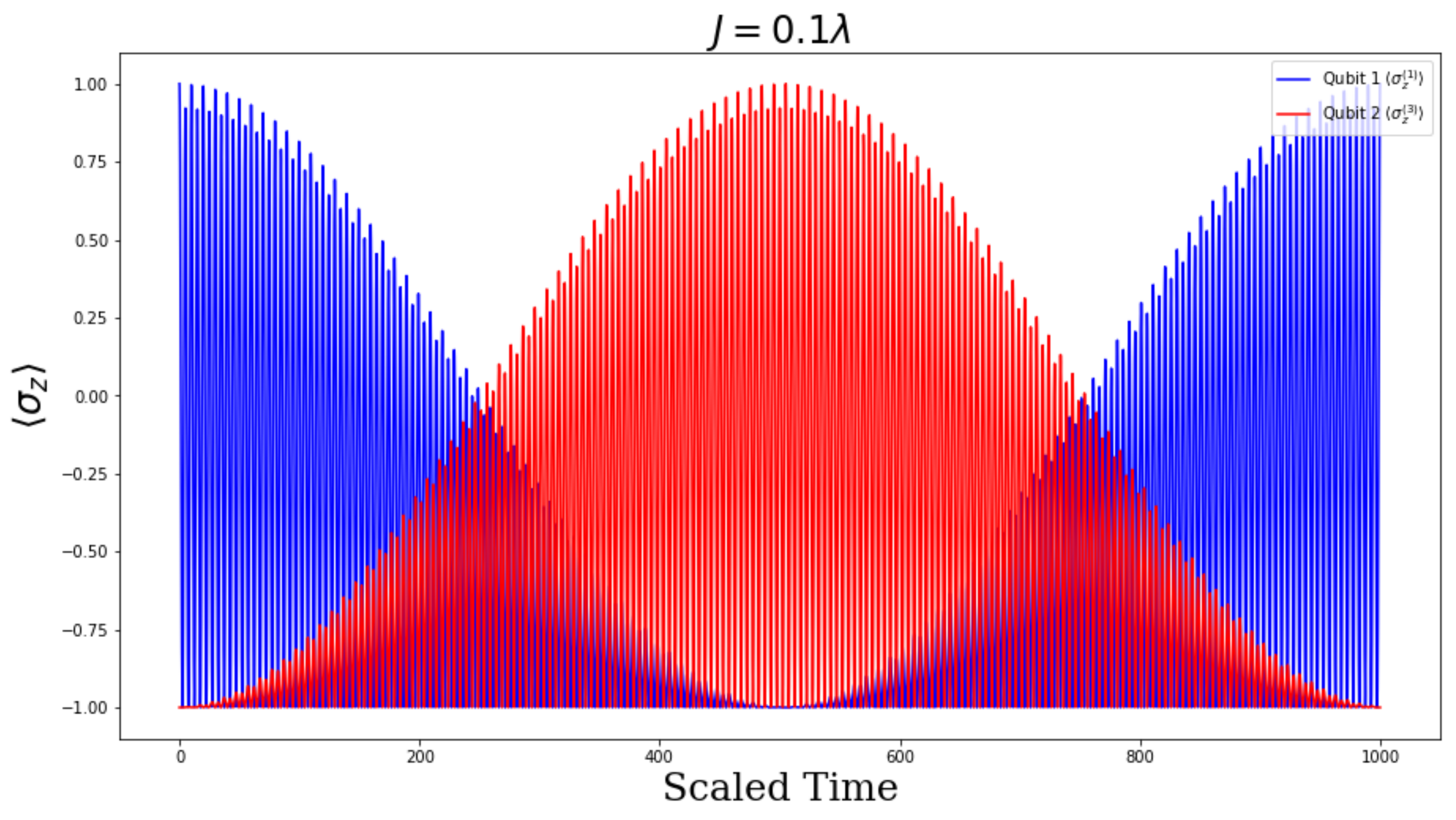}}
	\subfigure[]{\label{figure3_3b}\includegraphics[width=0.45\textwidth]{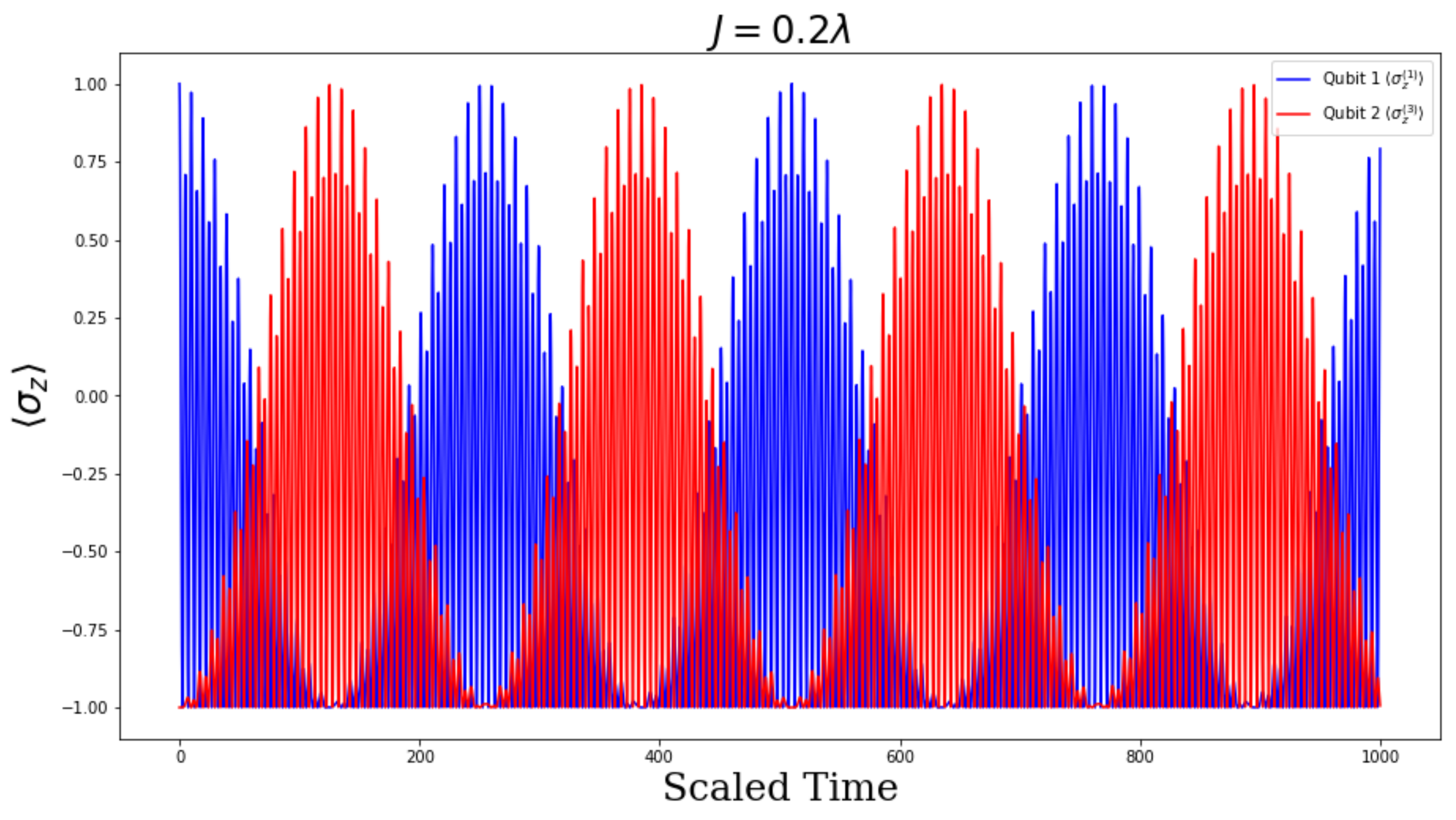}}
	\caption{Population inversion of qubit 1 and qubit 2 for (a)$J_{lm}=0.1\lambda_i$ and (b)$J_{lm}=0.2\lambda_i$.}
\end{figure}

\begin{figure}
	\centering
	\includegraphics[width=\linewidth]{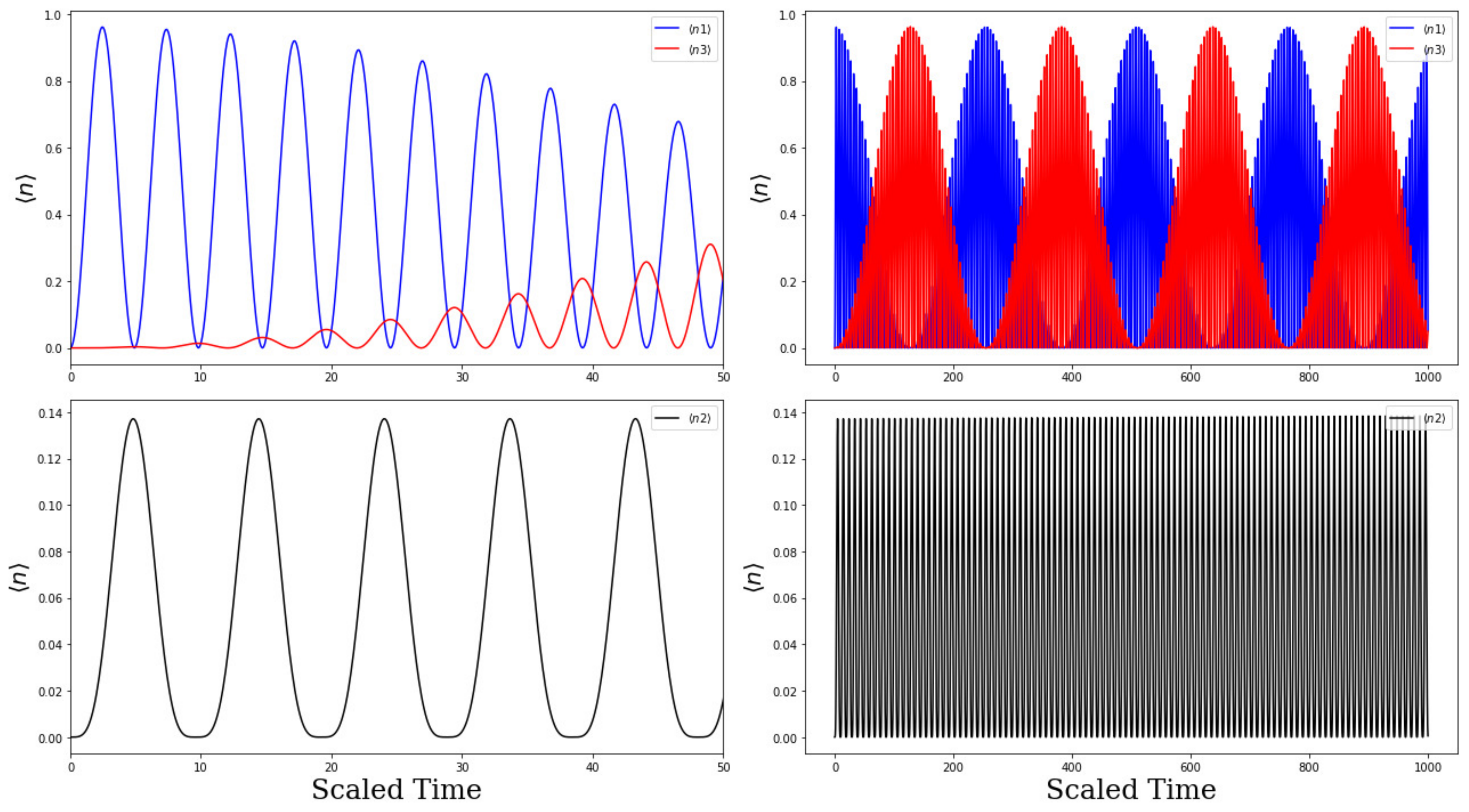}
	\caption{Photon number in 3 cavities $\braket{n_1}$, $\braket{n_2}$ and $\braket{n_3}$ for $J_{lm}=0.2\lambda_i$.The range of time differs in the plots on two columns.}
	\label{figure4}
\end{figure}

First we consider quantum state transfer for different coupling schemes, without an intermediate Kerr medium and  we can see that, there is a transfer of state from the qubit 1 to qubit 2 . This time can be controlled by controlling the coupling parameter. With $J_{lm}=0.1\lambda_i$ and $J_{lm}=0.2\lambda_i$, the results are shown in figures (\ref{figure3_3a}) and (\ref{figure3_3b}). The effect of coupling scheme is evident from these simulations. The photon number, $\braket{n}$ in each cavities are also estimated for $J_{lm}=0.2\lambda_i$. Figure (\ref{figure4}) shows the corresponding results. Here we have taken $\omega_a=\omega_c=\omega=2\pi f$ and $f$ is taken to be $1$GHz for the simulations. The graphs are plotted against scaled time. Here in all the plotting the unit of scaled time is in nanoseconds. All other parameters are scaled with respect to the unit of $\omega$.

The photon number also follows a pattern similar to the population inversion, such that  whenever  the qubit is in the excited state, photon number becomes zero inside the  respective cavity. However, the intermediate cavity shows a rise in the photon number as the coupling in the system is increased so that, the photon number in the first and last cavity has reduced. This suggest that the probability of inversion is not 100\% in any case with an intermediate coupling cavity with a non zero cavity-cavity coupling. So the expense of a controlled transmission with an intermediate coupling cavity is the quality of the transmission. The value of qubit-cavity coupling is kept at,  $\lambda=0.1\omega$, which allows us to use RWA \cite{Tur2000}.

\begin{figure}
	\centering
	\includegraphics[width=\linewidth]{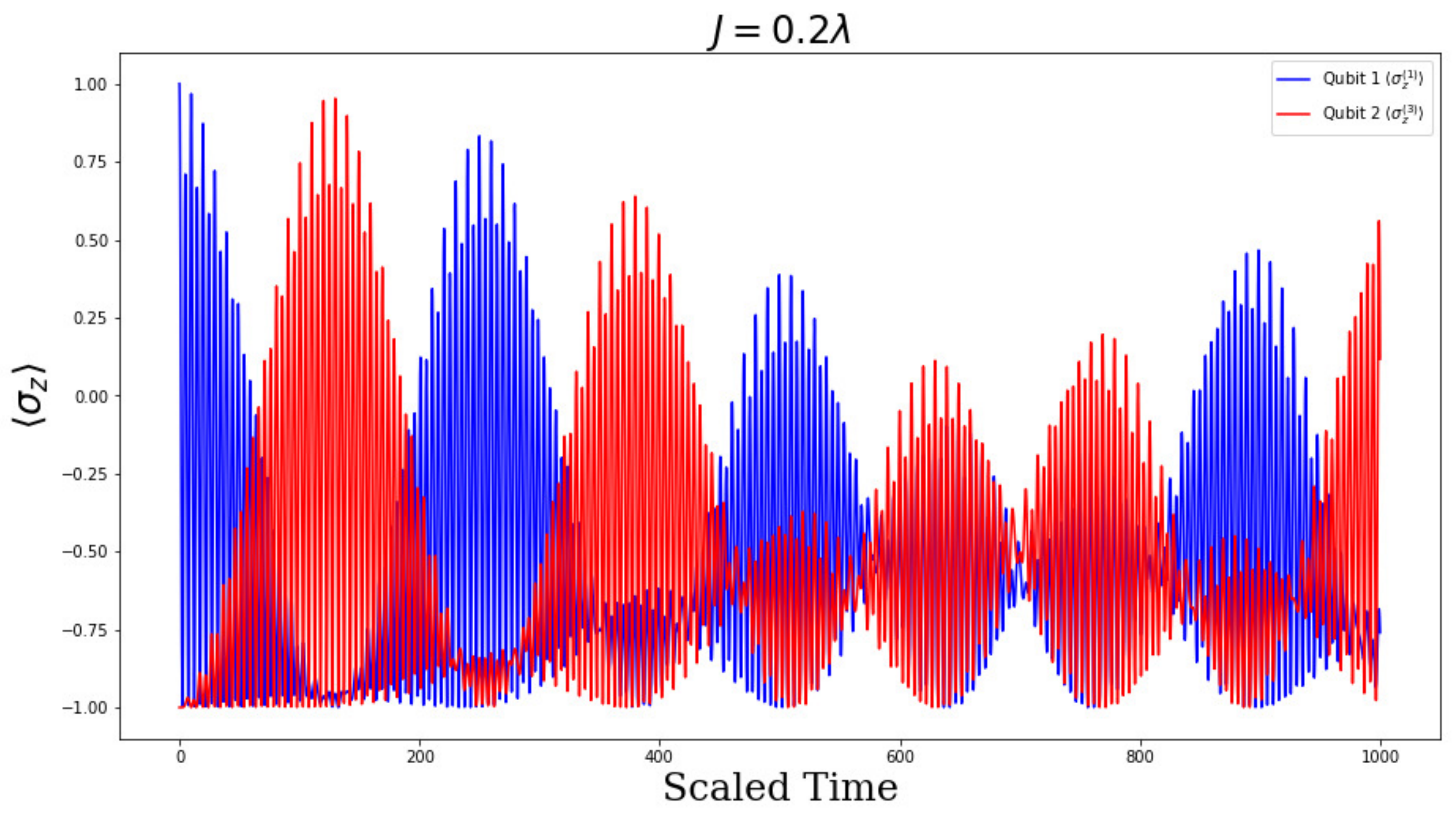}
	\caption{Population inversion of qubit 1 and qubit 2 for $J_{lm}=0.2\lambda_i$ and $\omega_c=0.99 \omega_a$ ($\Delta=0.01\omega_a$)}
	\label{figure5}
\end{figure}

\begin{figure}
	\centering
	\includegraphics[width=\linewidth]{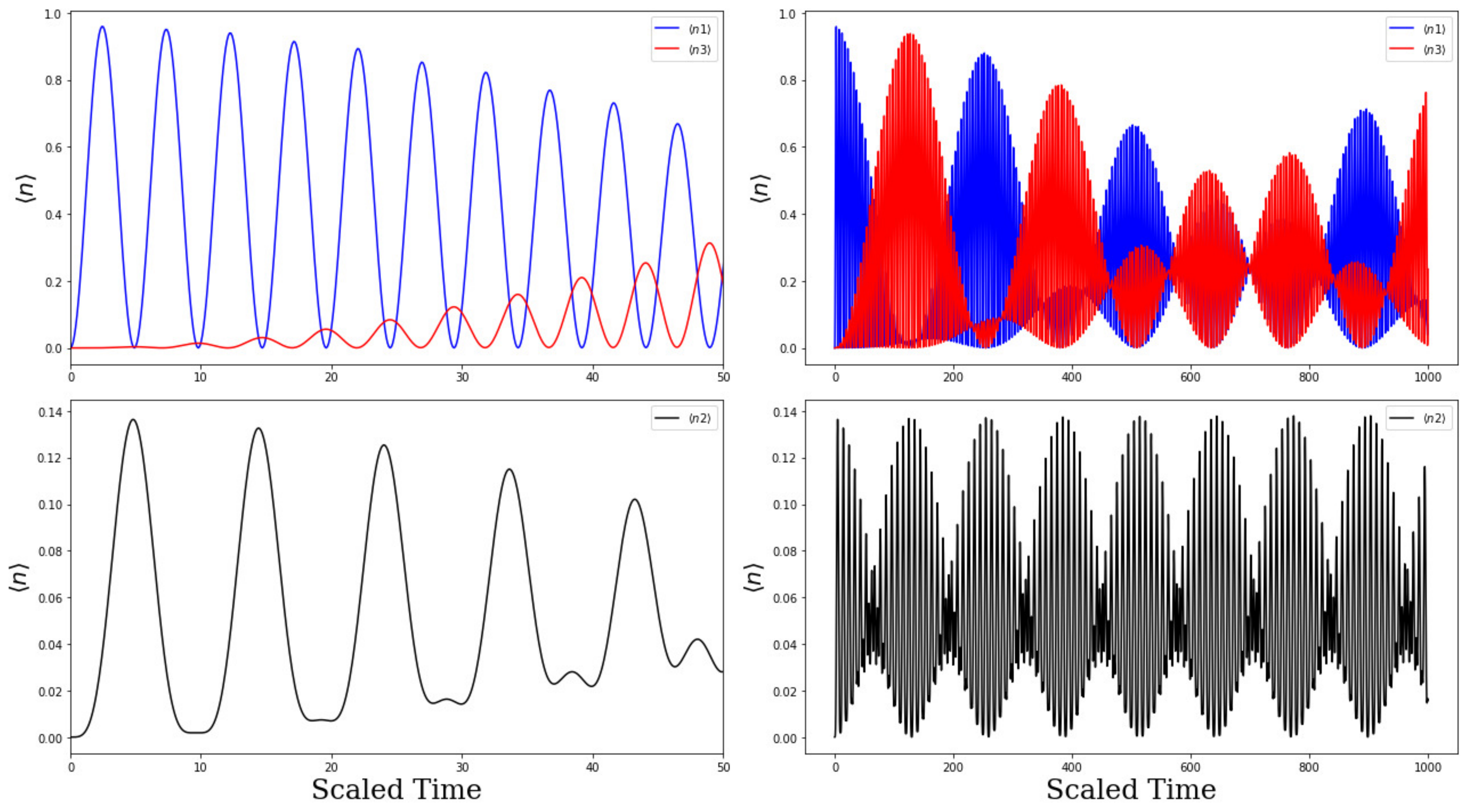}
	\caption{Photon number in 3 cavities $\braket{n_1}$, $\braket{n_2}$ and $\braket{n_3}$ for $J_{lm}=0.2\lambda_i$ and $\omega_c=0.99 \omega_a$($\Delta=0.01\omega_a$). The range of time differs in the plots on two columns.}
	\label{figure6}
\end{figure}

\noindent The analytical solutions including detuning is very cumbersome. The effect of detuning on the state transfer is shown in figures (\ref{figure5}) and (\ref{figure6}). We can clearly see that detuning affects the state transfer.  

\subsection{Quantum state transfer with Kerr medium}    

\begin{figure}
	\centering     
	\subfigure[]{\label{figure7_7a}\includegraphics[width=0.45\textwidth]{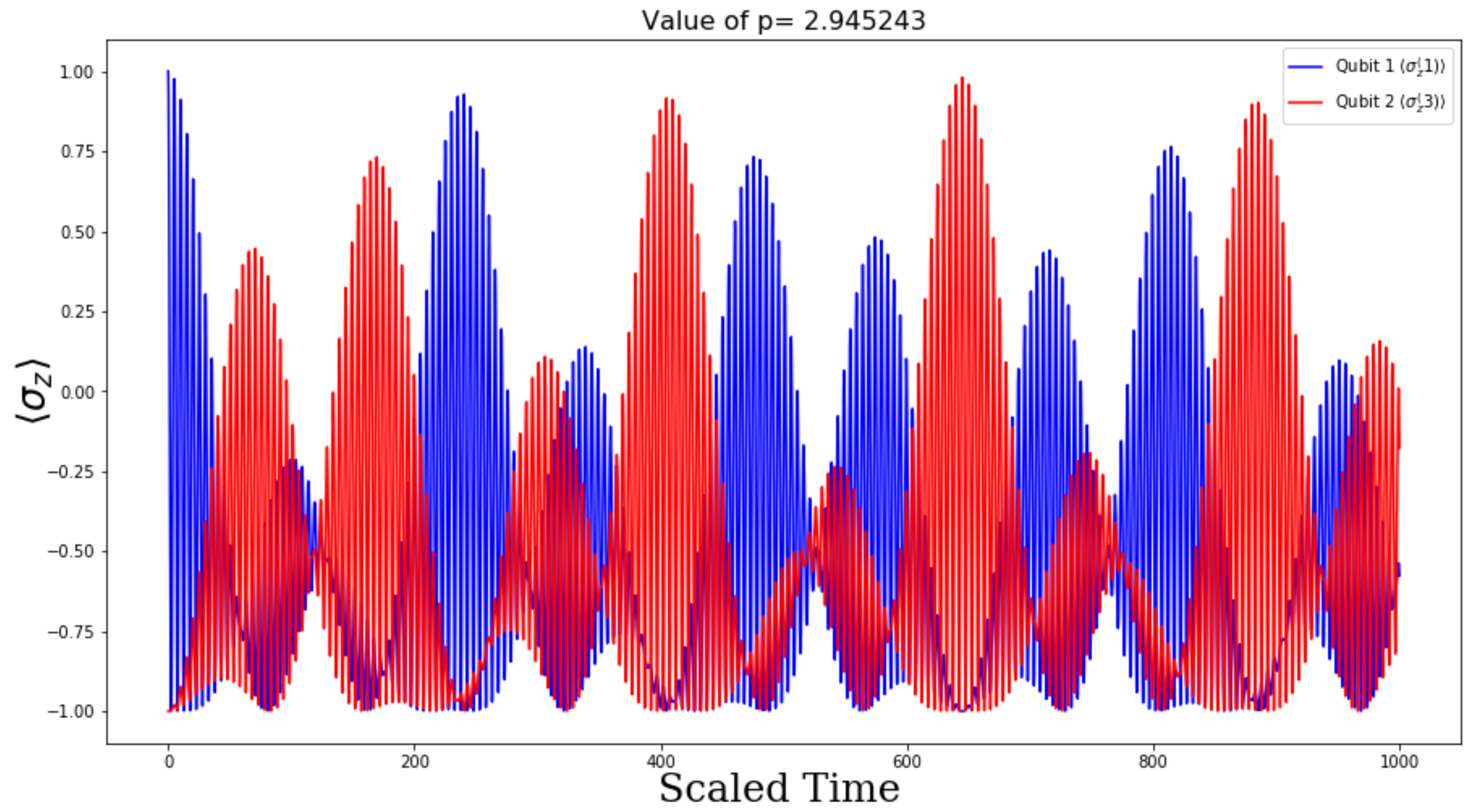}}
	\subfigure[]{\label{figure7_7b}\includegraphics[width=0.45\textwidth]{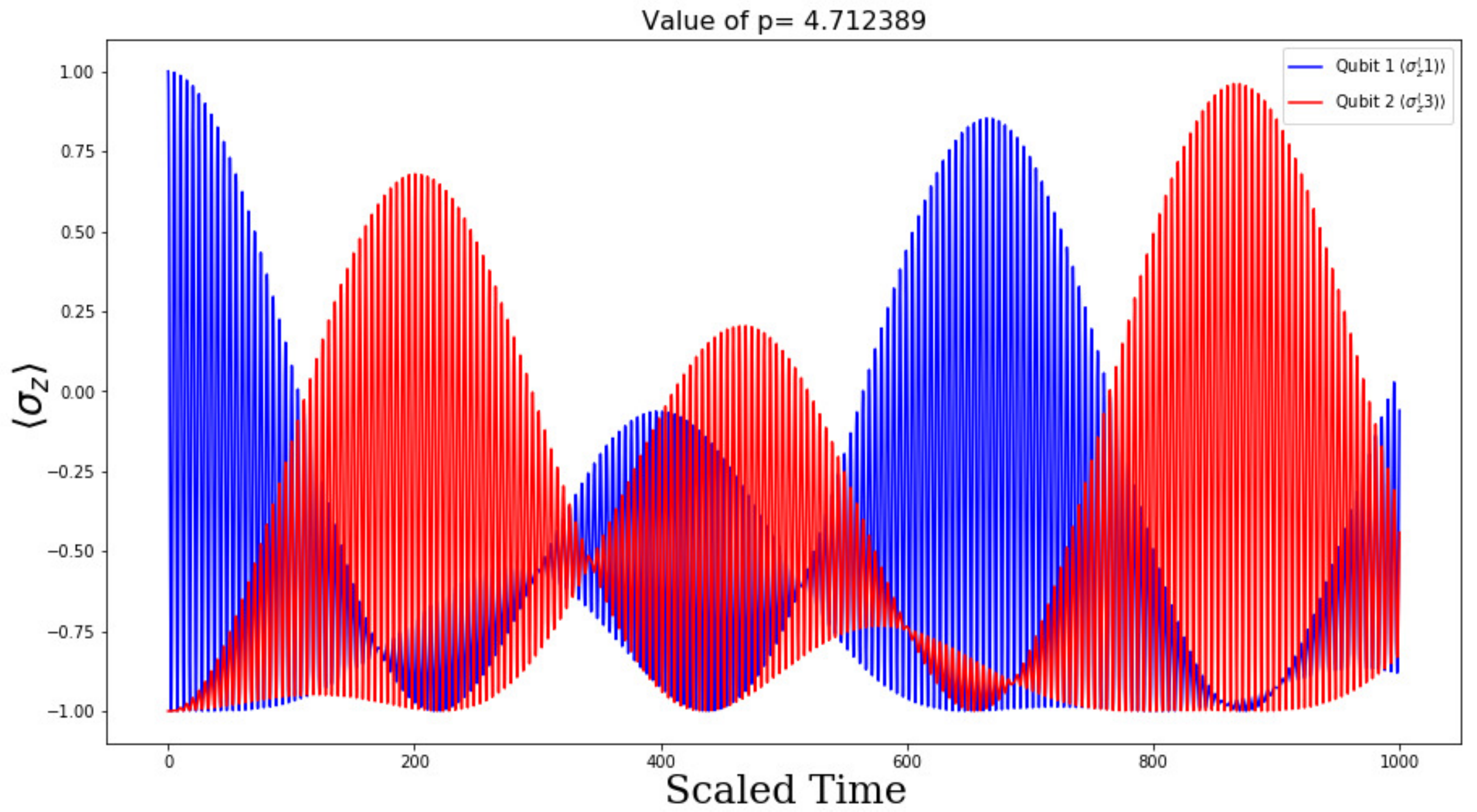}}
	\subfigure[]{\label{figure7_7c}\includegraphics[width=0.45\textwidth]{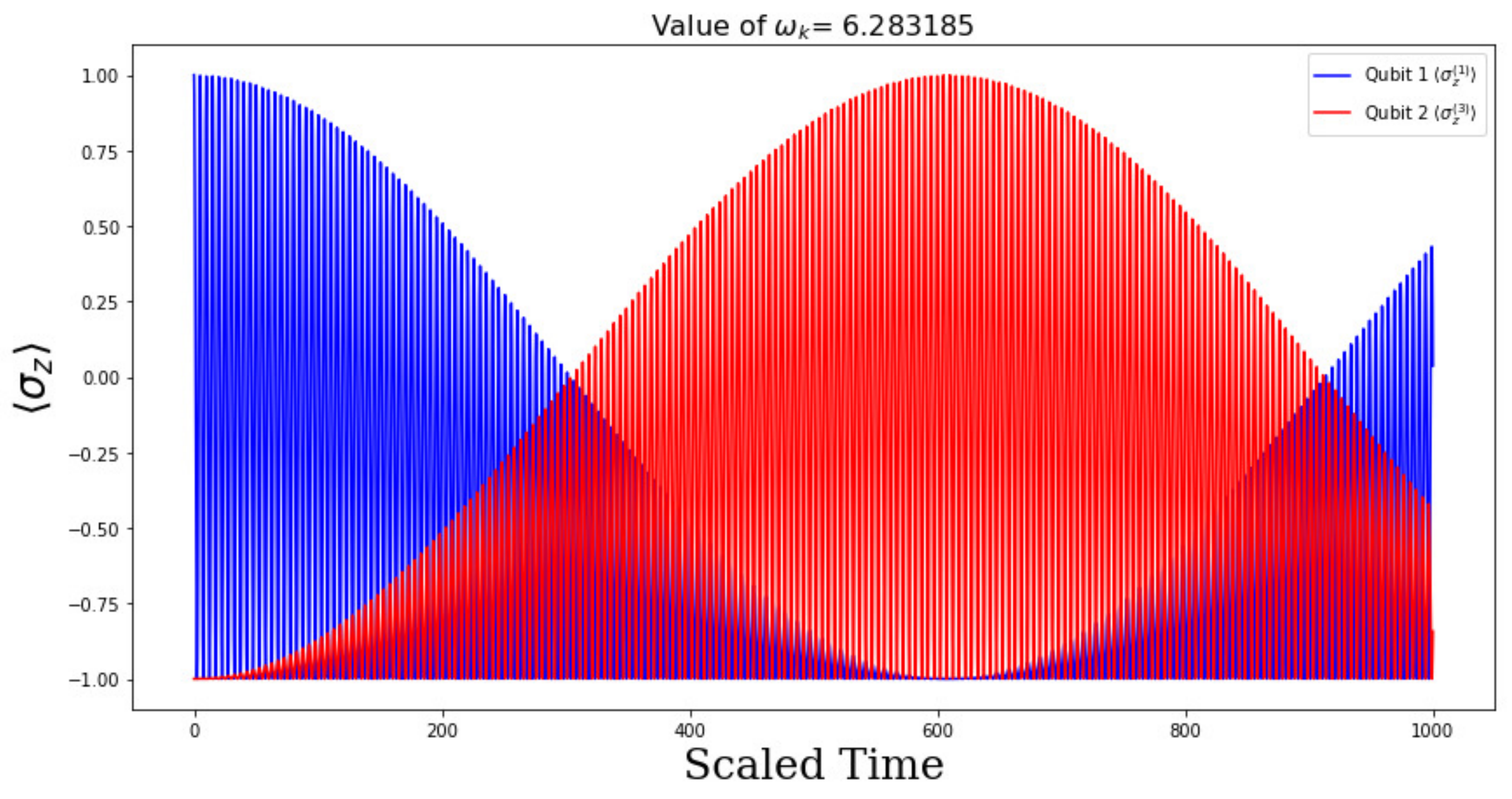}}
	\subfigure[]{\label{figure7_7d}\includegraphics[width=0.45\textwidth]{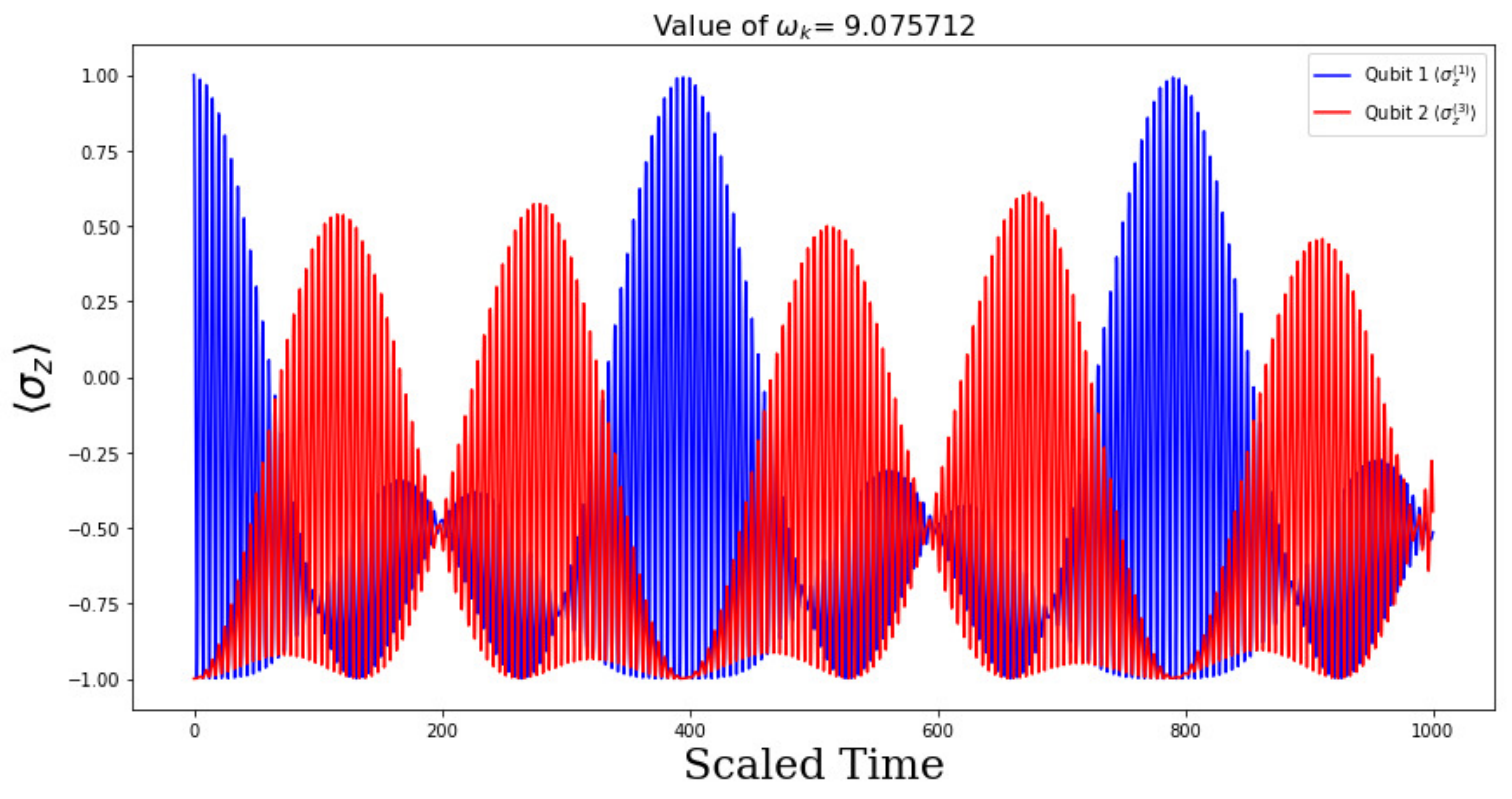}}
	\caption{(a) and (b) $\braket{\sigma_z}$ of the system under different $p$. Here $J=0.5\lambda_i$, $\lambda_i=0.1\omega_c$, $\omega_k=0.5\omega_c$ and $q=0.2\omega_c$, (c) and (d) $\braket{\sigma_z}$ of the system under different $\omega_{{K}}$. Here $J=0.5\lambda_i$, $\lambda_i=0.1\omega_c$, $p=0.557\omega$ and $q=0.2\omega_c$.}
\end{figure}

The presence of a Kerr medium in the intermediate cavity can affect the state transfer. Numerical simulations are done for different Kerr-cavity coupling value, $p$ and $\omega_{{K}}$. The results are shown in figures (\ref{figure7_7a}), (\ref{figure7_7b}), (\ref{figure7_7c}) and (\ref{figure7_7d}). Here we do not take the adiabatic approximation for the Kerr Hamiltonian.

\begin{figure}
	\centering     
	\subfigure[]{\label{figure8_8a}\includegraphics[width=0.45\textwidth]{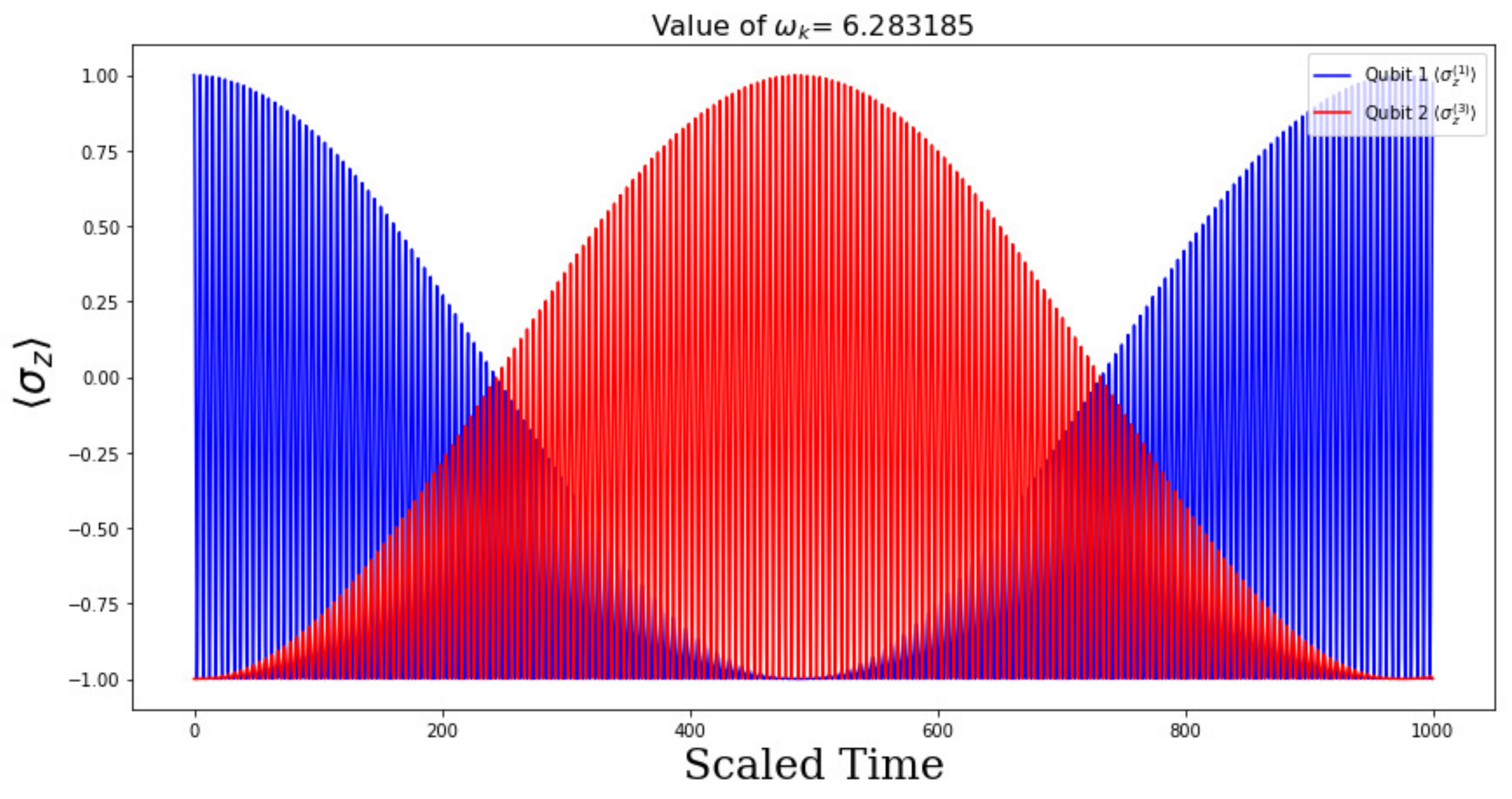}}
	\subfigure[]{\label{figure8_8b}\includegraphics[width=0.45\textwidth]{figure3_3a}}
	\caption{Population inversion of qubit 1 and qubit 2 for (a) $J_{lm}=0.5\lambda_i$, $p=0.5\omega_c$, $q=0.2\omega_c$ and $\omega_k=\omega_c$, with Kerr medium in the second cavity and (b) $J=0.1\lambda_i$, $\lambda_i=0.1\omega_c$ without Kerr medium in the second cavity}
\end{figure}

For $p\approx0.5 \omega_c$, $\omega_{{K}}\approx\omega_c$ and $J_{lm}=0.5\lambda_i$, the nature of population inversion is almost equivalent to the case where there is no Kerr medium in the second cavity and $J_{lm}=0.1\lambda_i$. The results are shown in figures (\ref{figure8_8a}) and (\ref{figure8_8b})

\noindent Thus with higher coupling between the cavities, we can have a controlled state transfer between two qubits, by means of a Kerr medium in the intermediate cavity.


\newpage
\section{Conclusion}
In the present work we have numerically studied a system of 3 linearly coupled cavities with one qubit in either ends of the cavity and an intermediate cavity in between them. Our study focused on the presence of a Kerr medium in the second cavity and how it affect the quantum state transfer. We found  that the presence of Kerr medium can affect the transmission and hence can be used as a quantum state transfer controller in quantum information processing. Without taking the adiabatic approximation in the Kerr medium, there can be a controlled state transfer. All the plotting are done with a scaling corresponds to the cavity frequency, which set at $1$ GHz. 
 We have only taken the the RWA in the appropriate limit.


\section*{Acknowledgments}
The authors, MTM and RBT would like to thank the financial support from KSCSTE, Government of Kerala State, under Emeritus Scientist scheme.

\bibliographystyle{ieeetr}
\bibliography{reference}

\end{document}